\documentclass[runningheads]{llncs}
\usepackage[T1]{fontenc}
\usepackage{subcaption}
\usepackage{graphicx}
\usepackage{xurl}
\usepackage{xspace}

\usepackage{microtype}
\newcommand{\mysubsection}[1]{\vspace{0.10cm}\noindent\textbf{#1}}

\newcommand{\toolname}{Epico\xspace}

\begin{document}
\title{Epico: Long-Lived WebAssembly Components for High-Performance Serverless Stream Processing}
\titlerunning{Epico: Serverless Stream Processing via Wasm}
%
\author{Matteo Della Bartola\inst{1}\orcidID{0009-0009-8415-2522} \and
Valerio Besozzi\inst{1}\orcidID{0009-0002-8493-2122} \and
Patrizio Dazzi\inst{1}\orcidID{0000-0001-8504-1503} \and Marco Danelutto\inst{1}\orcidID{0000-0002-7433-376X}} 
\authorrunning{M. Della Bartola et al.}
%
\institute{Dept. of Computer Science, University of Pisa, Pisa, Italy \\ 
\email{matteo.dellabartola@phd.unipi.it}, \email{valerio.besozzi@phd.unipi.it}, \email{patrizio.dazzi@unipi.it}, \email{marco.danelutto@unipi.it}}
\maketitle              
\begin{abstract}
While serverless computing is popular, its dominant Function-as-a-Service (FaaS) model is ill-suited for stream processing because its stateless, centrally orchestrated functions cannot efficiently handle continuous, low-latency event flows. We introduce \textit{\toolname}, a serverless runtime explicitly designed to resolve these inefficiencies at the runtime level. \toolname executes pipeline stages as persistent WebAssembly components, enabling independent, zero-to-infinity autoscaling based on queue-depth SLOs and routing events directly between stages using broker-free ZeroMQ channels. To optimize short execution paths, it utilizes a credit-based sliding window to amortize inter-process communication costs. Evaluations demonstrate that Ahead-of-Time (AOT) compilation reduces cold-start latencies from hundreds of milliseconds to sub-millisecond ranges, while the credit window improves single-worker throughput by up to \(4.3\times\). Compared to Apache OpenWhisk, \toolname bypasses the orchestrator bottlenecks and container overheads that typically hinder FaaS streaming workloads.

\end{abstract}
\section{Introduction}
Serverless computing is a cloud execution model that allows developers to deploy fine-grained, automatically scale applications without managing operational logic. Function-as-a-Service (FaaS) is the most widely adopted serverless model, characterized by event-driven execution and stateless, ephemeral functions that scale elastically. However, this model introduces significant overheads, particularly cold starts, which inflates latency when functions must be initialized from scratch.

These architectural limitations become especially problematic when running continuous stream-processing pipelines on general-purpose serverless platforms. Framework like Apache OpenWhisk does not provide true pipeline parallelism; instead, they operate stages as distinct invocations coordinated sequentially by a central controller. This introduces an orchestrator bottleneck, as intermediate data must be routed through the controller before triggering subsequent stages.

To address these challenges, we present \toolname, a serverless stream-processing runtime that executes pipelines defined as linear Directed Acyclic Graphs (DAGs). By replacing traditional containers with WebAssembly (Wasm) and the Webassembly System Interface (WASI), \toolname achieves lightweight sandboxing with microseconds-level instantiation times. Our framework introduces three architectural contributions: (1) \textbf{per-stage autoscaling}, where each pipeline stage scales automatically and independently from zero to a configured maximum based on its input queue depth; (2) \textbf{credit-based flow control}, which maximizes throughput using a TCP-inspired sliding window protocol to amortize round-trip costs between dispatchers and workers; and (3) \textbf{optimized execution} supporting Ahead-of-Time (AOT) compilation, which is critical for maintaining predictable, low-latency cold starts in production-grade environments.

\section{Background}
\label{sec:background}
\mysubsection{Serverless.} Serverless computing is a cloud computing execution model that allows users to deploy and execute \textit{fine-grained billed} and \textit{automatically scaled} applications, without having to address the underlying operational logic~\cite{11250936}.
Depending on the level of abstraction, three main serverless service models can be identified: \textit{Function-as-a-Service (FaaS)}, \textit{Backend-as-a-Service (BaaS)}, and \textit{Container-as-a-Service (CaaS)}. Serverless computing can be positioned within the cloud computing hierarchy alongside \textit{Platform-as-a-Service (PaaS)}. However, serverless is distinguished by its event-driven invocation model, automatic scaling mechanisms, and emphasis on stateless and ephemeral services.

%

\mysubsection{Function-as-a-Service (FaaS).} FaaS represents the most widely adopted form of serverless computing. In this service model, developers decompose applications into fine-grained, event-driven functions that are automatically deployed, executed, and scaled by the cloud provider. FaaS is characterized by the \textit{separation of computation from storage} and an \textit{event-driven} execution model~\cite{10.1145/3368454}, as functions are typically designed to be stateless and ephemeral, enabling elastic scaling from zero to virtually unlimited instances on demand. While this model enables elasticity and fine-grained resource utilization, it also introduces overheads, most notably the \emph{cold start}, which impacts latency when functions are invoked after periods of inactivity due to the time required to initialize and configure a new execution environment from scratch before executing the triggered function.

\mysubsection{WebAssembly.} 
Recent research has explored the use of WebAssembly~\cite{WebAssemblyCoreSpecification1} as lightweight execution environment for serverless platforms. WebAssembly provides language-level sandboxing with fast startup times and wide compatibility across programming languages. Approaches aim to reduce startup overheads and improve resource efficiency compared to traditional containerization. In addition, on top of the core WebAssembly specification, the \textit{WebAssembly System Interface (WASI)}~\cite{WASI} provides a system-level interface that exposes  modules that enable WebAssembly applications to interact with the underlying operating system. The WASI extension allows WebAssembly to be used not only for web-based applications.
It is important to note that the WebAssembly specification defines a binary instruction format and execution model for a stack-based virtual machine, but does not provide a reference runtime implementation. Consequently, several runtimes implementing the WebAssembly specification have emerged.
Among these, \textit{Wasmtime}~\cite{wasmtime}, developed by the Bytecode Alliance, represents one of the most widely adopted standalone runtimes. It provides secure, isolated execution and full support for WASI, enabling functions to interact safely with the host operating system. Furthermore, it is designed to achieve microsecond-level instantiation times, making it particularly suitable for mitigating cold-start overheads in serverless environments.

\section{\toolname}
\begin{figure}[h!]
\centering
\includegraphics[width=0.98\textwidth]{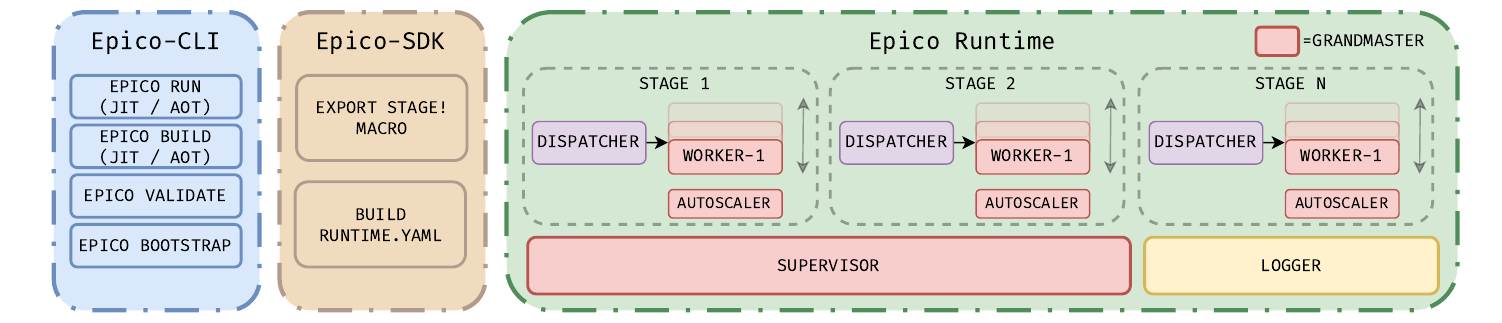}
\caption{Proposed architecture of Epico}
\label{fig:diagram_arch}
\end{figure}
\mysubsection{Overall Architecture}

\toolname\footnote{Sourcecode available at https://github.com/Della97/Epico}  is a serverless stream-processing runtime in which the developer defines a pipeline as a linear Directed Acyclic Graph (DAG) of stages. Each stage is defined using Webassembly that exports a single typed entry point, \texttt{process-event(ev, bench) -> (ev', bench')}, and scales automatically and independently from zero to a per-stage maximum based on its input queue depth.

The runtime consists of three primary actors (Fig. \ref{fig:diagram_arch}): a \textit{GrandMaster}, which owns the shared Wasmtime engine, supervises all dispatcher sub-processes, and coordinates the full lifecycle of a pipeline run, from loading and validating WebAssembly components to writing the summary telemetry upon shutdown, a \textit{Dispatcher} for inter-stage communication, and a \textit{Logger} for logging purposes.

For each pipeline stage, the GrandMaster spawns a dedicated \textit{autoscaler} thread that ticks every $1\,\mathrm{ms}$. The autoscaler polls its corresponding stage's dispatcher for queue depth via a ZMQ REQ/REP control socket and drives a vote-with-cooldown controller. Each tick contributes one vote in the direction implied by the current queue depth: a scale-up vote when queue depth exceeds an upper threshold, a scale-down vote when it falls below a lower threshold. The autoscaler commits a single-replica scaling action only after the corresponding counter crosses a configurable threshold, and adjusts capacity by one replica at a time.
When scaling up, the autoscaler spawns a \textit{worker}—an OS thread that holds one live Wasmtime component instance for its entire lifetime. Each worker connects two ZMQ sockets: a \textit{DEALER} socket to its own stage's dispatcher (from which it pulls events) and a \textit{PUSH} socket to the subsequent stage's dispatcher.

The \textit{Worker} loop deserializes each incoming JSON event into a \texttt{Val::Record}, calls the component's \texttt{process-event} export, serializes the output back to JSON, appends a per-hop timing entry to the event's benchmark context, and signals readiness back to the dispatcher. When scaling down, the autoscaler sets a drain flag on the target worker; the worker then exits cleanly at the next loop iteration.

The \textit{Supervisor} module manages the dispatcher child processes. It spawns them at startup in a downstream-first order, installs a Ctrl+C handler, and gracefully kills the entire process tree upon shutdown.

The Supervisor spawns one \textit{Dispatcher} process for each stage. The Dispatcher is a single-threaded ZeroMQ broker exposing three sockets: a \textit{PULL} frontend to receive events from the previous stage, a \textit{ROUTER} backend to which replicas connect with \textit{DEALER} sockets, and a \textit{REP} control socket the autoscaler queries via \textit{REQ}. 
The Dispatcher's primary role is to accept events from the producer, distribute them to ready replicas under a worker-pull dispatch strategy: only replicas that have signaled readiness are eligible to receive events, so the ROUTER pushes events in FIFO order over the ready queue. This is structurally least-loaded without explicit load measurement.
Internally, the Dispatcher maintains three data structures. First, an \textit{event buffer}: a FIFO of pending events received on the frontend. Once it reaches the configured maximum, the dispatcher stops polling the frontend, which propagates backpressure upstream via the PULL socket's high-water mark rather than dropping events.
Second, a \textit{ready queue}: a FIFO of replica identities representing free dispatch slots, with one entry per credit held: a replica with $K$ credits appears $K$ times. The queue and the per-replica credit count are kept in sync as an invariant: each successful dispatch pops one entry and decrements that replica's credit count by one; each credit refill pushes $K$ entries and increments credits by $K$.
Third, a \textit{worker registry}: a hash map keyed by replica identity recording its remaining credit count, its lifetime dispatched-event count, and its most recently reported metric payload. The registry is consulted on every successful send and on every control reply.


Replica departure is detected lazily and handled uniformly across causes. When the autoscaler drains a replica, the worker thread exits at its next loop iteration but sends no notification to the dispatcher. The dispatcher therefore has no proactive liveness mechanism and discovers the departure only when its next \texttt{send\_multipart} to that identity returns \texttt{EHOSTUNREACH}. This mechanism handles graceful drains and unexpected disappearances indistinguishably, and triggers a bulk cleanup that removes the identity from the registry and evicts all of its remaining credit entries from the ready queue in one $\mathcal{O}(n)$ pass.

\mysubsection{Credit-Based Flow Control.}
A strict request-reply protocol between dispatcher and worker, equivalent to a credit window of 1, bottlenecks per-worker throughput on the dispatcher to worker round trip, irrespective of how cheap the WebAssembly stage is. To break this bottleneck, \toolname implements a credit-based sliding window protocol inspired by TCP flow control, layered on top of the existing DEALER-ROUTER socket pair between worker and dispatcher.
On boot, each worker sends a \textit{hello} message advertising a credit window $N$; the dispatcher records this as the worker's initial credit balance. The dispatcher is then free to send up to $N$ events to that worker without waiting, decrementing the worker's credit on each send. The worker processes events as they arrive, once it has completed at least $N/2$ events, sends a single \textit{refill} message carrying both the latest execution telemetry and the number of credits to replenish. The dispatcher blocks only when all workers in the pool have exhausted their credits.
The window amortizes the round-trip between dispatcher and worker across $N/2$ events at the cost of a tradeoff: backpressure becomes per-window rather than per-event. Delivery is currently best-effort: the event that triggers \texttt{EHOSTUNREACH} is retained and redispatched, but events already in flight to the departed worker are lost. Achieving at-least-once delivery would require a per-worker "recently sent" buffer on the dispatcher that replays unacknowledged events.

\mysubsection{Workflow to Instantiate a Pipeline.}
Using the proposed framework, instantiating a pipeline requires only two developer actions and a single CLI invocation. First, the developer writes a \texttt{pipeline.yaml} file describing the record types, stages (with their input/output types, source paths, and scaling parameters), and the \texttt{edges} defining the linear DAG. Second, they provide one Rust file per stage containing the business logic wrapped in the \texttt{stage!} macro.
Running \texttt{\toolname build} then parses the YAML configuration and generates the \texttt{target/\toolname/} workspace, creating one crate per stage with the corresponding \texttt{wit/world.wit}, \texttt{src/lib.rs}, and \texttt{runtime.yaml}. Each crate is compiled into a WASI-P2 component (\texttt{.wasm}, optionally \texttt{.cwasm}). Finally, \texttt{\toolname run} starts the runtime by launching the GrandMaster, which spawns a dispatcher for each stage.



\section{Experimental Evaluation}
We evaluate \toolname along three axes: ablation studies that isolate the impact of individual architectural choices, a comparison between serverless and serverfull deployment with Epico, and a head-to-head comparison against OpenWhisk as a general-purpose serverless platform.

\mysubsection{Hardware.} All benchmark tests were executed on a single x86 machine equipped with an AMD Ryzen 7 5700U processor (8 cores @ 1.8 GHz) and 16 GB of RAM.

\mysubsection{Workloads.} We implemented two stream-processing pipelines designed to ingest synthetically generated IoT sensor data. The \textit{Simple Implementation} imports no external Rust modules, establishing a baseline for execution overhead. The \textit{Complex Implementation} imports various external libraries, accurately reflecting the heavier compilation logic of a production-grade application.

\mysubsection{Baselines and Configurations.} For our ablation studies, we evaluated our framework natively by toggling compilation strategies (AOT vs. JIT~\cite{colosi2025serverlesseverywherecomparativeanalysis}) and assessing the impact of our credit-based flow control mechanism. For the state-of-the-art comparison, we deployed logically equivalent pipelines in OpenWhisk. We utilized the composition operational mode to mimic a linear pipeline, with each stage executing the exact same Rust code used in our native framework.

\subsection{Ablation Study}

\mysubsection{Compilation Strategy.}
To evaluate the performance gains introduced by our framework's optimizations, we first isolate the impact of AOT compilation against a baseline JIT compilation. The primary objective is to assess reductions in cold-start latency across both the Simple and Complex implementations.

\begin{figure}[h!]
  \centering
  \begin{minipage}{0.99\textwidth}
    
    \begin{subfigure}{0.48\linewidth}
      \centering
      \includegraphics[width=\linewidth]{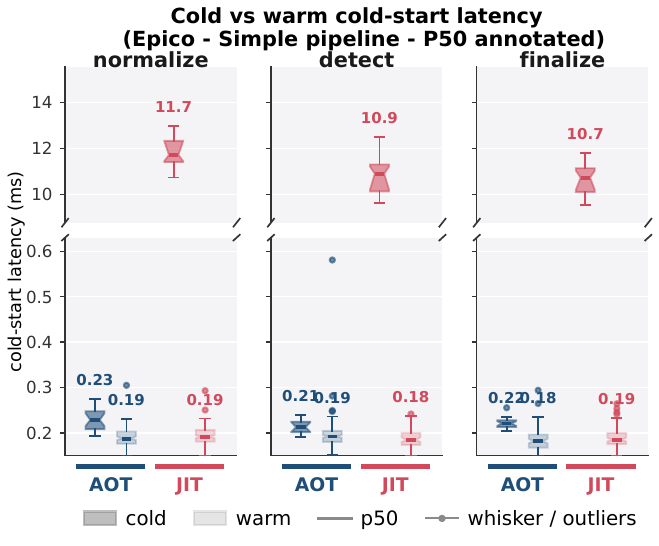}
      \caption{Simple Implementation Cold Start.}
      \label{fig:03_simple}
    \end{subfigure}\hfill
    \begin{subfigure}{0.48\linewidth}
      \centering
      \includegraphics[width=\linewidth]{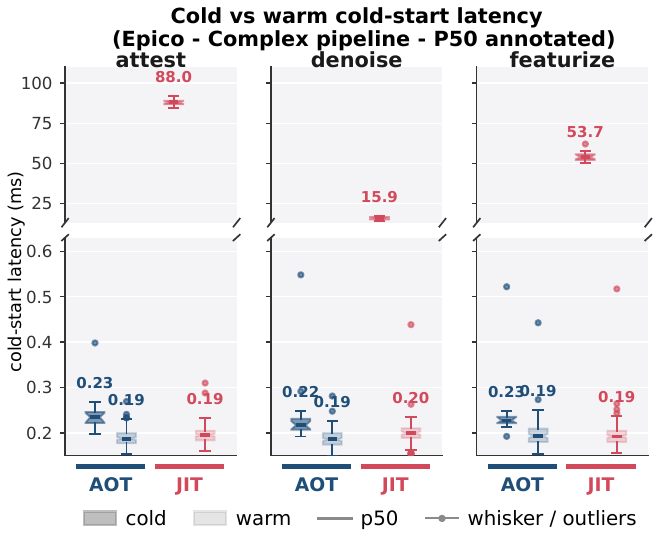}
      \caption{Complex Implementation Cold Start.}
      \label{fig:03_complex}
    \end{subfigure}
    
    \caption{Cold-start latency comparisons for Compilation Strategies over 20 AOT and 20 JIT runs (The y-axis is split to improve readability).}
    \label{fig:compilation_ablation_1}
    
  \end{minipage}
\end{figure}

Our results demonstrate that while JIT and AOT deliver identical steady-state performance, JIT introduces a severe latency penalty during cold starts that scales with application complexity. In the Simple Implementation, which imports no external modules, JIT compilation adds approximately 11 ms of overhead per stage (Fig.~\ref{fig:03_simple}). Consequently, the maximum end-to-end latency for the JIT pipeline reaches 42 ms, whereas the AOT one only 18 ms (Fig.~\ref{fig:fullvsless}).
The performance gap widens drastically in the Complex case. The inclusion of external dependencies raises the JIT compilation time, resulting in first-invocation latencies of 88.0 ms, 15.9 ms, and 53.7 ms across the three pipeline stages (Fig.~\ref{fig:03_complex}).

\begin{figure}[h!]
    \centering
    \includegraphics[width=0.99\linewidth]{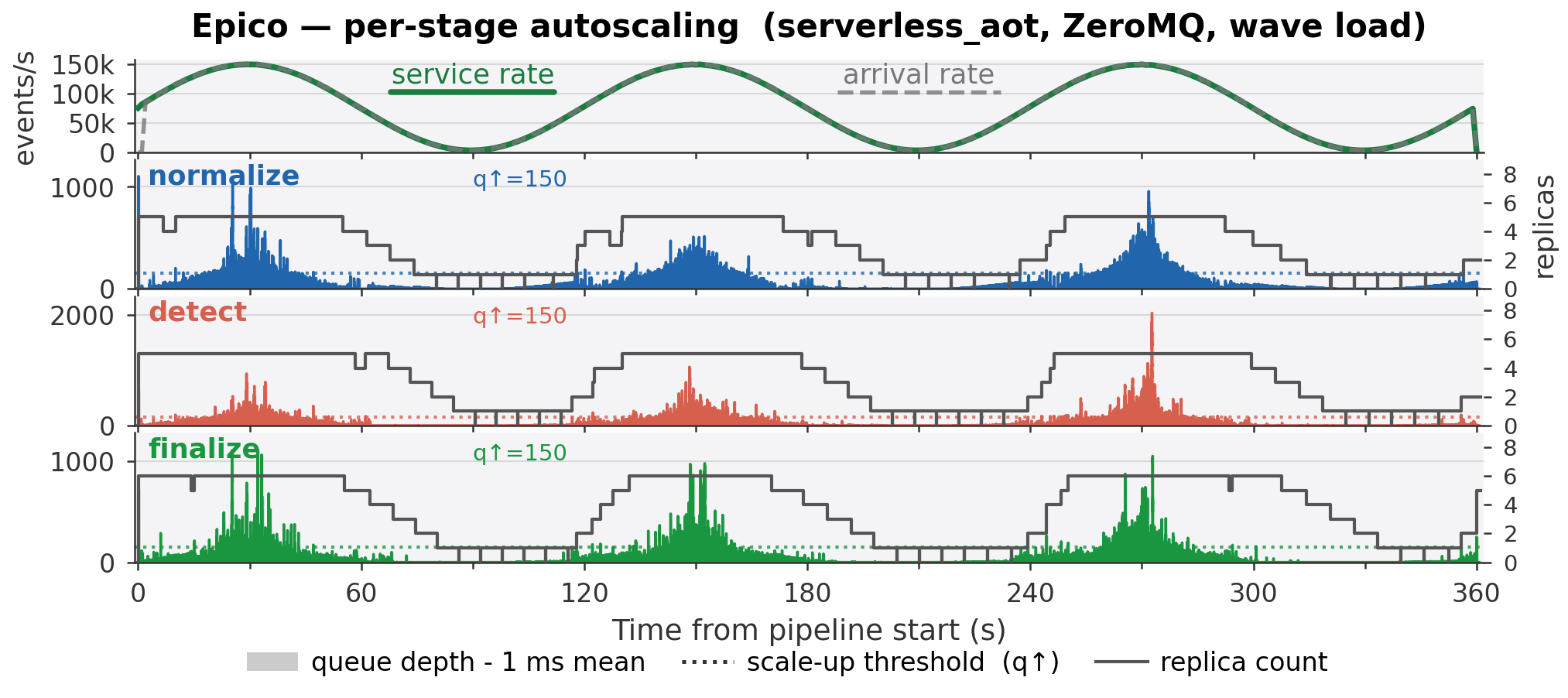}
    \caption{Per stage independent scaling.}
    \label{fig:scaling_ind}
\end{figure}

\begin{figure}[h!]
\centering
\includegraphics[width=0.80\textwidth]{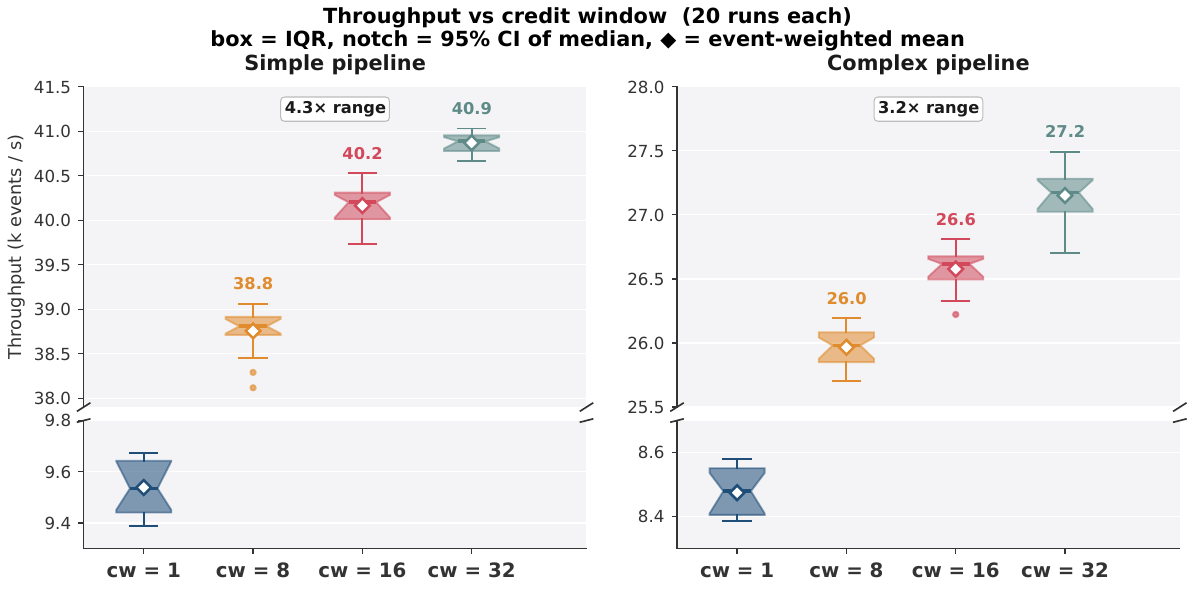}
\caption{Throughput comparison across different credit window sizes for the Simple and Complex pipelines. (20 runs each credit window).} 
\label{fig:window}
\end{figure}

\mysubsection{Credit-Based Flow Control.}
We next evaluate the throughput gains of our credit-based flow control protocol (Fig.\ref{fig:window}). For these tests, both the Simple and Complex applications were deployed as linear three-stage pipelines constrained to exactly one worker max replica per stage. In a strict request-reply baseline (a credit window of 1), synchronous telemetry replies bottleneck the dispatcher-to-worker IPC. This caps single-worker throughput at 9.7 kev/s for the Simple pipeline and 8.5 kev/s for the Complex pipeline.
By implementing a sliding window protocol that permits up to $N$ unacknowledged events in flight, we successfully amortize this round-trip cost. Toggling the credit window to 8 yields noticeable improvements: capacity jumps to 38.8 kev/s (a 4$\times$ increase) for the Simple pipeline and 26.0 kev/s (a 3.0$\times$ increase) for the Complex pipeline. 
Further increasing the window to 16 and 32 provides marginal gains, plateauing at around 40.0 kev/s and 27.0 kev/s respectively. Beyond $N=8$ the bottleneck shifts from the dispatcher-to-worker path to the dispatcher's single-threaded per-event work. This confirms that a credit window of 8 optimally hides IPC overhead, enabling high-throughput execution that is impossible under a strict request-reply model.

\subsection{Performance Comparison}

\begin{figure}[h!]
    \centering
    \includegraphics[width=0.85\linewidth]{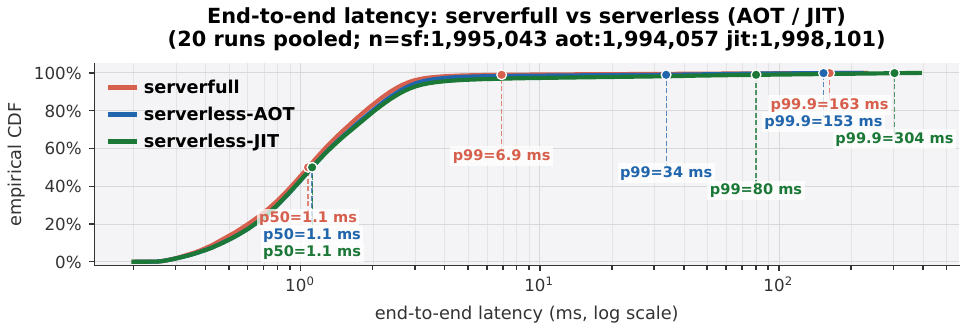}
    \caption{E2E latency serverfull vs serverless (AoT and JIT).}
    \label{fig:fullvsless}
\end{figure}

\mysubsection{Serverless vs. Serverfull.}
We compare our framework under a serverless scaling model, where each stage scales independently based on queue depth (Fig.~\ref{fig:scaling_ind}), against a serverfull deployment with a static replica allocation. Both configurations were tuned via grid search to achieve the same median end-to-end latency of 1.1~ms (Fig.~\ref{fig:fullvsless}). In the serverfull setup, replicas were fixed at their optimal capacities, whereas the serverless deployment dynamically scaled each stage up to a configured maximum. While delivering equivalent P50 latency, the serverless approach reduced resource consumption by $36.5\%$ in terms of replica-seconds, averaged over 20 runs. These results show that autonomous per-stage autoscaling reclaims idle compute capacity during load troughs without sacrificing typical pipeline latency.

\begin{figure}[h!]
  \centering
  \begin{minipage}{0.99\textwidth}
    
    \begin{subfigure}[b]{0.45\linewidth}
      \centering
      \includegraphics[width=\linewidth]{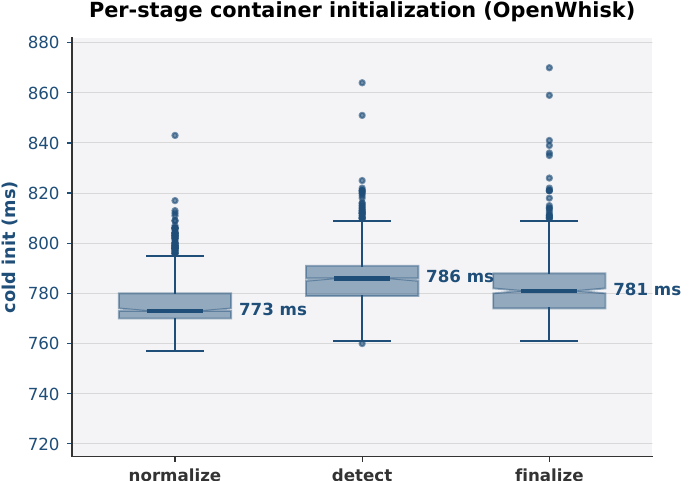}
      \caption{Per-stage init latency.}
      \label{fig:ow-latency}
    \end{subfigure}\hfill 
    \begin{subfigure}[b]{0.45\linewidth}
      \centering
      \includegraphics[width=\linewidth]{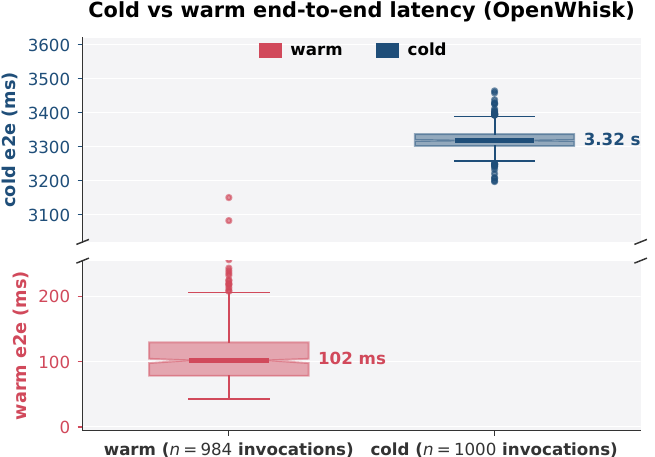}
      \caption{Cold vs warm E2E latency.}
      \label{fig:ow-e2e}
    \end{subfigure}
    
    \caption{Performance evaluation of the Simple pipeline in OpenWhisk.} 
    \label{fig:ow-performance}
    
  \end{minipage}
\end{figure}

\mysubsection{OpenWhisk Comparison.}
We compared our native stream-processing framework against Apache OpenWhisk. While OpenWhisk's composition mode achieves platform-level concurrency, it does not provide true pipeline parallelism. Stages do not maintain continuous execution; instead, they operate as distinct serverless invocations sequentially coordinated by a central controller.
This architecture inherently suffers from an orchestrator bottleneck. Intermediate data must be routed back through the controller before triggering the next stage, preventing direct, high-throughput handoffs. Furthermore, OpenWhisk relies on containerization, which introduces severe cold-start penalties. As observed in our tests, each individual stage incurs nearly 800 ms of initialization during a cold start: 773 ms for normalization, 786 ms for detection, and 781 ms for finalization (Fig.~\ref{fig:ow-latency}).
Cumulatively, this drives the E2E cold-start latency to 3.32 s (Fig.~\ref{fig:ow-e2e}). Even when fully warm, OpenWhisk's centralized routing overhead inflates the pipeline's end-to-end latency to 102 ms, resulting in a 32.5$\times$ gap between cold and warm states.
Beyond latency, OpenWhisk scales these stateless invocations blindly against a global namespace limit. Without per-stage capacity controls, a single slow stage can indiscriminately spawn containers, saturate system resources, and throttle the entire workflow. In contrast, our framework eliminates these structural flaws by utilizing Wasmtime for microsecond-level instantiation, ZeroMQ for direct, broker-free handoffs, and granular limits for stage replication.

\section{Related Works}
\mysubsection{WebAssembly for Serverless.} 
Several works investigate the use of WebAssembly as a lightweight execution environment for serverless functions~\cite{9214403,hall_execution_2019,inproceedingspushing,besozzi2025webassemblyunikernelscomparativestudy}.
Kjorveziroski et al.~\cite{Kjorveziroski2023WebAssemblyAA} analyze the performance of popular WebAssembly runtimes in terms of cold start delay and execution time, comparing JIT and AOT compilation. 
Other works focus on custom runtimes that leverage WebAssembly for serverless computing at the edge.
Hall et al.~\cite{hall_execution_2019} present a WebAssembly-based serverless runtime for edge environments built on top of Node.js and V8, showing lower cold-start latency compared to container-based deployments.
Similarly, addressing serverless at the edge, Gadepalli et al.~\cite{gadepalli_challenges_2019,10.1145/3423211.3425680} propose \textit{aWsm} and \textit{Sledge}. The former is a WebAssembly runtime that targets edge devices. The latter is a serverless framework built on top of aWsm. It leverages AoT compilation to LLVM to enable low-latency execution and efficient concurrency management on resource-constrained devices.
For compute-intensive and distributed workloads, Shillaker et al.~\cite{shillaker_faasm_2020,305991} introduce \textit{Faasm}, a distributed serverless runtime based on WebAssembly. Faasm provides lightweight isolation through \textit{Faaslets}, enabling memory sharing between functions while reducing cold-start overheads through a snapshot-based mechanism.
While these works demonstrate the potential of WebAssembly for reducing cold-start overheads and improving execution isolation in serverless environments, they primarily focus on general-purpose FaaS platforms. In contrast, our work investigates WebAssembly for a stream processing serverless runtime, emphasizing low-latency stage-to-stage communication, per-stage autoscaling, and high-throughput workflow execution.

\mysubsection{Workflow-Oriented Serverless Frameworks.}
Recent works focus on serverless frameworks for orchestrating workflows and DAG-based applications across cloud, edge, and HPC infrastructures.
HyperFlow~\cite{MALAWSKI2020502}, UniFaaS~\cite{10579158}, OSCAR~\cite{8814513},  and SCAR~\cite{PEREZ201850} focus on scientific workflow orchestration and federated execution across heterogeneous infrastructures, including cloud platforms, accelerators, and HPC systems.
Other systems target performance-oriented DAG execution. WUKONG~\cite{10.1145/3419111.3421286} employs decentralized scheduling and task clustering to improve data locality and reduce coordination overheads for burst-parallel workloads on AWS Lambda. Similarly, DataFlower~\cite{10.1145/3623278.3624755} adopts a data-flow execution model to minimize orchestration overheads and improve responsiveness in serverless workflows.
In contrast to these works, \toolname{} targets stream processing applications rather than workflow orchestration, and a single host rather than a federated infrastructure. Stages are long-lived consumers of a continuous event stream, autoscaled per stage against queue-depth SLOs, rather than coarse-grained tasks scheduled per invocation by a DAG engine. The current prototype deliberately restricts topologies to linear DAGs and defers fan-out, fan-in to future works. The focus is instead on the runtime mechanics that make wasm stages viable as a serverless execution targeting stream processing.

\mysubsection{Serverless Stream Processing.}
To the best of our knowledge, few works specifically target serverless stream processing runtimes. \textit{Sponge}~\cite{288745} uses serverless functions to elastically absorb workload bursts in stream-processing pipelines and scale on-demand in case of peak load.  However, it primarily acts as an auxiliary elasticity layer for existing stream processing application, whereas our work provides a standalone serverless runtime with native stage-level autoscaling.
In contrast, \textit{SPSC}~\cite{10568243}  is conceptually more similar to our work and offers a serverless framework built on top of public FaaS platforms. Our work differentiates from this by not relying on existing general-purpose serverless platform by providing a serverless runtime that manages directly stage scaling, inter-stage (inter-function) communication, and by using WebAssembly as an execution environment.

\section{Future Works}

Several components of the current prototype represent natural directions for future work.

\mysubsection{Autoscaling.} The current autoscaler relies solely on dispatcher queue depth. Extending it with latency, downstream backpressure, and host resource metrics would improve decisions under diverse workloads.

\mysubsection{Topology.} The prototype currently supports only linear DAGs. Adding fan-in and fan-out would broaden the range of supported stream-processing applications.

\mysubsection{Distributed execution.} While distributed deployment is already possible through manual configuration, future work includes automatic stage placement, transparent transport selection, and host-aware autoscaling.

\mysubsection{Dynamic workflow adaptation.} The execution graph is currently static. Supporting runtime DAG transformations (e.g., stage fusion and scission) and policy- or LLM-driven sub-workflow creation would enable adaptive execution while preserving application semantics.

\section{Conclusions}

We presented \toolname, a serverless stream-processing runtime in which pipelines are long-lived WebAssembly components, scaled independently per stage from zero against queue-depth SLOs, and connected through direct broker-free ZeroMQ channels. We showed that AOT reduces cold-start latency from tens or hundreds of milliseconds to sub-millisecond, and that a credit-based sliding-window protocol between workers and dispatchers raises per-worker throughput by up to $4.2\times$ relative to a strict request–reply baseline. Compared head-to-head with Apache OpenWhisk on logically equivalent pipelines, \toolname avoids the orchestrator bottleneck and the container-driven cold-start penalty that bound general-purpose FaaS platforms when used for streaming workloads.






\begin{credits}
\subsubsection{\ackname} This work has been partially funded by the NOUS (A catalyst for EuropeaN ClOUd Services in the era of data spaces, high-performance and edge computing) HORIZON-CL4-2023-DATA-01-02 project, G.A. n. 101135927, and by Spoke 1 "FutureHPC \& BigData" of the Italian Research Center on High-Performance Computing, Big Data and Quantum Computing (ICSC) funded by MUR Missione 4 Componente 2 Investimento 1.4: Potenziamento strutture di ricerca e creazione di "campioni nazionali di R\&S (M4C2-19 )" - Next Generation EU (NGEU).
\end{credits}
%
\bibliographystyle{splncs04}
\sloppy
\bibliography{bib}
\end{document}